\documentclass[prd,preprint,tightenlines,floatfix,showpacs,preprintnumbers,nofootinbib,eqsecnum]{revtex4-1}

\pdfoutput=1

\usepackage[dvips,final]{graphicx}
\usepackage{color}
\usepackage{amssymb}
\usepackage{amsmath}
\usepackage{amsfonts}
\usepackage{epsfig}
\usepackage{bm}
\usepackage{comment}
\usepackage{float}
\usepackage[utf8]{inputenc}

\def\GeV{\,\mbox{GeV}}

\def\dfr{\mathrm{d}}

\def\Jpsi{J\!/\!\psi}   

\def\Y{\Upsilon}

\DeclareSymbolFont{myletters}{OML}{ztmcm}{m}{it}
\DeclareMathSymbol{\uplambda}{\mathord}{myletters}{"15}

\begin{document}

\title{Exclusive photoproduction of excited quarkonia\\ 
in ultraperipheral collisions}

\author{Cheryl Henkels$^{1}$}
\email{cherylhenkels@hotmail.com}

\author{Emmanuel G. de Oliveira$^{1}$}
\email{emmanuel.de.oliveira@ufsc.br}

\author{Roman Pasechnik$^{1,2,3}$}
\email{Roman.Pasechnik@thep.lu.se}

\author{Haimon Trebien$^{1}$}
\email{haimontrebien@outlook.com}

\affiliation{
\\
{$^1$\sl Departamento de F\'isica, CFM, Universidade Federal 
de Santa Catarina, C.P. 476, CEP 88.040-900, Florian\'opolis, 
SC, Brazil
}\\
{$^2$\sl
Department of Astronomy and Theoretical Physics, Lund
University, SE-223 62 Lund, Sweden
}\\
{$^3$\sl Nuclear Physics Institute ASCR, 25068 \v{R}e\v{z}, 
Czech Republic\vspace{1.0cm}
}}

\begin{abstract}
\vspace{0.5cm}

In this paper, we discuss the exclusive photoproduction of ground and excited 
states of $\psi(1S,2S)$ and $\Upsilon(1S,2S)$ in ultraperipheral collisions (UPCs). 
Using the potential model in order to obtain the vector meson wave function, we find 
a good agreement of our calculations with data from the LHC and HERA colliders 
for $J/\psi (1S,2S) $ and $\Upsilon(1S)$ in $\gamma p$ collisions. We extend 
the calculations to the nuclear target case applying them to $AA$ UPCs with the use of 
the shadowing and finite coherence length effects fitted to the data. Our results 
are compared to the recent LHC data, in both incoherent ($J/\Psi(1S)$ at 2.76 TeV) 
and coherent ($J/\Psi(1S)$ at 2.76 and 5.02 TeV) processes. We also show 
the corresponding predictions for the excited states, in the hope that future 
measurements could provide more detailed information about the vector meson 
wave functions and nuclear effects.

\end{abstract}

\pacs{14.40.Pq,13.60.Le,13.60.-r}

\maketitle

\section{Introduction}
\label{Sect:intro}

Phenomenology of exclusive quarkonia photoproduction processes offers very sensitive and 
powerful probes for the associated soft and hard QCD phenomena. In the case of bottomonia
$\Y(1S,2S)$ photoproduction, the heavy quark mass $m_Q$ provides a sufficiently hard scale 
for the perturbative QCD framework to be applicable for a precise description of the 
underlying production mechanism \cite{Kopeliovich:1991pu,Kopeliovich:1993pw,Nemchik:1994fp,Nemchik:1996cw}.
However, charmonia $\psi(1S,2S)$ photoproduction probes predominantly nonperturbative
phenomena at a semihard scale. This means that a simultaneous description of the existing
charmonia and bottomonia photoproduction data is required for validation of the universality 
of the quarkonia production mechanism that is expected to incorporate both hard and soft QCD
effects on the same footing. For a detailed review on quarkonia physics, see e.g.~Refs.~\cite{Ivanov:2004ax,Brambilla:2010cs} 
and references therein.

A notable progress in understanding of the mechanisms of heavy quarkonia elastic electro- 
(with large photon virtuality $Q^2\gg 0$) and photo- (with quasireal photon $Q^2=0$) production 
has been done over the past two decades starting from $ep$ collisions at HERA 
\cite{Adloff:2000vm,Alexa:2013xxa,Breitweg:1998ki,Chekanov:2002xi,Chekanov:2009zz}. More 
recently, the ultra-peripheral $pA$ and $AA$ collisions (UPCs) at the LHC have provided 
clean experimental means for probing the real photoproduction mechanisms of heavy quarkonia 
in photon-Pomeron fusion with intact colliding nucleons or nuclei, thanks to low QCD 
backgrounds. Over the past few years, a wealth of phenomenological information on elastic 
(exclusive) and quasi-elastic $J/\psi\equiv \psi(1S)$ and $\psi'\equiv \psi(2S)$ 
photoproduction in UPCs has become available from the LHC measurements, in particular, 
from LHCb \cite{Aaij:2015kea,LHCb:2018ofh,Aaij:2018arx}, ALICE \cite{Abelev:2012ba,Abbas:2013oua,Adam:2015sia,Kryshen:2017jfz,Acharya:2019vlb} 
and CMS \cite{Khachatryan:2016qhq,Sirunyan:2018sav} experiments. 
Meanwhile, the existing theoretical approaches remain rather uncertain due to poorly 
known nonperturbative and a $D$-wave admixture in the corresponding $S$-wave 
quarkonia wave functions \cite{Krelina:2018hmt,Cepila:2019skb,Krelina:2019egg}, as well as 
the coherence phenomena particularly relevant for photoproduction in $AA$ UPCs 
\cite{Kopeliovich:1991pu,Hufner:2000jb,Kopeliovich:2001xj,Ivanov:2002kc}.

A standard view on quarkonia production mechanism is encapsulated in the framework of non-relativistic 
QCD (NRQCD) where one assumes a small relative intrinsic motion of the heavy (non-relativistic) 
quark and antiquark, with the perfectly harmonic interaction potential (see e.g.
Refs.~\cite{Frankfurt:1995jw,Nemchik:1996cw}). For charmonia photoproduction, nonperturbative 
and relativistic corrections may be significant since the mass of $c$-quark is not large enough 
in order to safely rely on the perturbative QCD approach. Besides, as of tradition, the lowest-order
transition amplitudes for an $S$-wave quarkonium $Q\bar Q \to V$ ($V=\psi(nS),\Y(nS)$, $n=1,2$) 
is conventionally assumed to have the same form as the lowest Fock state of the photon in 
the LF (or infinite-momentum) frame, $\gamma \to Q\bar Q$. Such an assumption about a 
photon-like LF quarkonium wave function is unjustified as the corresponding
rest-frame wave function necessarily has an admixture of a $D$-wave component whose weight cannot 
be established in the framework of a suitable interquark interaction potential \cite{Hufner:2000jb}. 
Such an uncontrollable $D$-wave contribution has a considerable impact on quarkonia
photoproduction observables as was recently advocated in Ref.~\cite{Krelina:2019egg}.

In this work, we perform an analysis of the exclusive quarkonia photoproduction in $AA$ 
UPC collisions at the LHC in the phenomenologically successful color dipole picture \cite{Kopeliovich:1981pz,Nikolaev:1994kk} 
(for an early analysis of diffractive charmonia photoproduction in the dipole framework, see
e.g.~Ref.~\cite{Kopeliovich:1991pu,Kopeliovich:1993pw,Nemchik:1994fp,Nemchik:1996cw,Ducati:2013bya}).
The main focus is the quarkonia photoproduction observables in $AA \to A+V+X$ in both the coherent
($X=A$) and incoherent ($X=A^*$ with $A^*$ being an excited state of the nucleus) scattering regimes.
Here, one of the important ingredients of the production amplitude is the Light-Front 
(LF) quarkonium wave function found in the framework of potential approach going beyond 
the NRQCD approximation. In order to avoid an unjustified $D$-wave effect one 
starts with a pure $S$-wave $Q\bar Q \to V$ transition in the $Q\bar Q$-pair rest frame as 
a product of spin-dependent and radial components. The radial component is found by the solution 
of the Schr\"odinger equation for a given model of the interquark interaction potential. 
A particularly relevant feature
of the excited quarkonia states is the presence of one or several nodes in their radial wave 
functions \cite{Nemchik:1996cw}. These imply a possible cancellation of contributions to the 
photoproduction amplitude coming from the regions below and above the node position, in particular,
in the $\psi'\equiv \psi(2S)$ wave function causing its relative suppression compared to 
the $\Jpsi$ production amplitude \cite{Nemchik:2000de}. In the spin-dependent component, 
a transformation of (anti)spinor of a heavy (anti)quark $Q$ $(\bar Q)$ from the $Q\bar Q$ 
rest frame to the LF frame known as the Melosh transform is employed. The resulting LF 
quarkonium wave function has been recently validated in a detailed analysis of the $S$-wave 
quarkonia electro- and photoproduction observables at HERA 
in Refs.~\cite{Krelina:2018hmt,Cepila:2019skb} providing a consistent estimate of the underlying 
theoretical uncertainties.

The article is organised as follows. In Sect.~\ref{sec:QQ-off-proton}, we provide a brief 
discussion of the LF potential approach for the proton target case and present the 
corresponding numerical results on the energy dependence of the $\gamma p \to V p$ 
integrated cross section compared against all the available data from the HERA, 
LHC and fixed-target experiments. In Sect.~\ref{sec:QQ-off-nucleus},
an analysis of coherent and incoherent quarkonia photoproduction off nuclear targets is performed
and the numerical results are shown against the available LHC data on $AA$ UPCs. Finally,
concluding remarks and a summary are given in Sect.~\ref{Sect:Concl}.

\section{Exclusive photoproduction off the proton target}
\label{sec:QQ-off-proton}

\subsection{$\gamma p\to V p$ cross section}
\label{sec:gammap-CS}

Consider first the case of exclusive quarkonia photoproduction in high-energy photon-proton 
$\gamma p\rightarrow V p$ scattering in the rest frame of the proton target. According to the
dipole picture \cite{Kopeliovich:1981pz,Nikolaev:1994kk}, in the high-energy regime, both
the photon and the heavy quarkonium $V=\psi(nS),\Y(nS)$ ($n=1,2$) can be considered
as color dipoles whose transverse separations are frozen during their interactions with
a target nucleon. The lowest-order $Q\bar Q$ Fock fluctuation of the projectile quasi-real 
photon scatters off the target nucleon at a certain impact parameter and with a fixed 
interquark separation. This is an elementary QCD process described through the universal 
dipole cross section as an eigenstate of the elastic scattering operator. The same $Q\bar Q$
dipole is then projected $Q\bar Q \to V$ to a given quarkonium state $V$ using 
the corresponding LF wavefunction. In the non-relativistic limit, the $Q\bar Q$ separation 
is found to be $r_V\simeq 6/M_V$, in terms of the quarkonium mass $M_V$ \cite{Nemchik:1994fp,Krelina:2018hmt}. The perturbative domain 
then corresponds to $r_V \lesssim r_g$ where the gluon propagation length 
in the nucleon $r_g\sim 0.3$ fm represents the soft scale of the process \cite{Kopeliovich:2004ex,Kopeliovich:2006bm}. In the case of photoproduction, 
this condition is satisfied only for bottomonia, while charmonia are produced 
predominantly in the soft regime.

The exclusive photoproduction cross section integrated over the impact parameter 
of $\gamma p$ collisions \cite{Hufner:2000jb}
\begin{eqnarray}
\sigma^{\gamma p\rightarrow Vp}(W) = \frac{1}{16\pi B}\,
\Big(\mathrm{Im}\mathcal{A}^{\gamma p\rightarrow V p}(W)\Big)^{2}  \,, 
\label{total-cs}
\end{eqnarray}
is found in terms of the forward exclusive photoproduction amplitude $\mathcal{A}(W)$,
the elastic slope parameter $B$ typically fitted to the exclusive 
quarkonia electroproduction data available from the HERA collider, and
$W$ is the $\gamma p$ center-of-mass energy. For the numerical values of the slope $B$ 
we adopt the most recent energy-dependent parametrisation of the HERA data known 
from Ref.~\cite{Cepila:2019skb}. 

The forward $Q\bar Q$ photoproduction amplitude reads
\begin{eqnarray}
\mathrm{Im}\mathcal{A}^{\gamma p\rightarrow V p}(W) =
\int\limits_0^1\dfr \beta \int\dfr^2r_\perp \,\Psi^{\dagger}_{V}(\beta,r_\perp)\,
\Psi_{\gamma}(\beta,r_\perp)\sigma_{q\bar q}(x,r_\perp) \,, \;\; 
x = \frac{M_V^2}{W^2} \,,
\label{psi-amp}
\end{eqnarray}
where $\Psi_{\gamma}(\beta,r_\perp)$ is the LF wave function of a transversely-polarized 
(quasi-real) photon fluctuating into a $Q\bar Q$ dipole and $\Psi_{V}(\beta,r_\perp)$ 
is the LF quarkonium wave function. The transverse $Q\bar Q$ 
dipole size is $\vec r_\perp$, $\beta=p_Q^+/p_{\gamma}^+$ is the longitudinal momentum 
fraction of the photon momentum $p_{\gamma}^+ = E_{\gamma} + p_{\gamma}$ carried 
away by a heavy quark $Q$. The universal dipole cross section describing the $Q\bar Q$ 
dipole elastic scattering off the proton target is $\sigma_{q\bar q}(x,r_\perp)$, 
where $x$ is the standard Bjorken variable used e.g. in diffractive DIS 
\cite{Ryskin:1995hz} and associated with the proton energy loss 
in the dipole-proton scattering. 

In the NRQCD limit, one typically assumes an equal energy sharing between $Q$ 
and $\bar Q$, such that the LF quarkonium wave function is approximated as 
$\Psi_{V}(\beta,r_\perp)\propto \delta(\beta-1/2)$ \cite{Kopeliovich:1991pu}. Another 
imposed approximation is a photon-like $Q\bar Q \to V$ transition amplitude, with the
radial wavefunction based upon a naive harmonic interquark potential
\cite{Frankfurt:1995jw,Nemchik:1996cw}. In our analysis below, we go beyond these 
approximations following the formalism of Ref.~\cite{Hufner:2000jb,Cepila:2019skb}
for the proton target, and then adopt it for the studies of quarkonia photoproduction 
observables in $AA$ UPCs as done in Refs.~\cite{Hufner:2000jb,Ivanov:2002kc}.

\subsection{Quarkonium wave function}
\label{sec:QQ-WFs}

A consistent computation of the LF quarkonium wave function $\Psi_{V}(\beta,r_\perp)$ 
in the infinite-momentum frame remains a challenging problem even for the lowest Fock 
$V\to |Q\bar Q\rangle$ state \cite{Hufner:2000jb}. In this work, we follow the potential 
approach of Ref.~\cite{Hufner:2000jb,Cepila:2019skb} starting from the factorised 
wave function for a pure $S$-wave state in the light-cone (LC) momentum 
representation defined as
\begin{eqnarray} 
\label{eq:PsiV-LF}
&& \Psi_{V}^{(\mu,\bar\mu)}(\beta,\vec p_T)=U^{(\mu,\bar\mu)}(\beta,\vec p_T)\psi_{V}(\beta,p_T) \,, 
\quad U^{(\mu,\bar\mu)}(\beta,\vec p_T)=\frac{1}{\sqrt{2}}\xi_Q^{\mu\dag}\vec\sigma
\vec e_{V}\tilde\xi_{\bar Q}^{\bar \mu} \,, \\
&& \tilde\xi_{\bar Q}^{\bar\mu}=i\sigma_y\xi_{\bar Q}^{\bar\mu\ast}\,, \qquad 
\xi^\mu_Q = R(\beta,\vec p_T)\chi_Q^\mu \,, \qquad 
\xi_{\bar Q}^{\bar\mu}=R(1-\beta,-\vec p_T)\chi_{\bar Q}^{\bar\mu} \,,
\nonumber
\label{eq:Psi_QQ}
\end{eqnarray}
in terms of the spin-dependent and spatial (radial) parts of the vector meson 
wave function denoted as $U^{(\mu,\bar\mu)}(\beta,\vec p_T)$ and $\psi_{V}(\beta,p_T)$,
respectively. Here, $\vec e_{V}$ is the vector 
meson polarisation vector, $\xi_Q^{\mu}$ and 
$\xi_{\bar Q}^{\bar\mu}$ are the heavy quark and antiquark spinors in the $Q\bar Q$ 
rest frame, respectively, related to their counterparts in the infinite momentum frame, 
$\chi_Q^{\mu}$ and $\chi_{\bar Q}^{\bar\mu}$, by means of the Melosh spin transformation 
matrix given by \cite{Melosh:1974cu,Hufner:2000jb}
\begin{eqnarray}
R(\beta,\vec p_T)=\frac{- i\vec p_T(\vec\sigma\times\vec n) + m_Q + \beta M_V}
{\sqrt{p_T^2+(m_Q+\beta M_V)^2}} \,.
\label{eq:Melosh}
\end{eqnarray}
One may assume that the corrections due to Melosh spin transformation are numerically 
not very relevant for a non-relativistic $Q\bar Q$ system. In the case of the 
ground-state quarkonium states such as $J/\psi$, indeed, the photoproduction 
cross section increases by roughly 30\% only. However, in the case of excited 
states the spin rotation effect is much larger and enhances $\psi(2S)$ photoproduction 
by about a factor 2--3 \cite{Hufner:2000jb,Krelina:2018hmt,Cepila:2019skb}.

The spatial wave function $\psi_{V}(\beta, p_T)$ is typically found by using 
a simple Lorentz boost prescription \cite{Terentev:1976jk} based upon 
the conservation of probability. Starting from the corresponding spatial 
wave function $\psi(p)$ in the $Q\bar Q$ rest frame one writes
\begin{eqnarray} 
\label{VMwave_boosting}
&& \psi_V(\beta,p_T) =
\left(\frac{p_T^2+m_Q^2}{16(\beta (1-\beta ))^3}\right)^{\frac{1}{4}}\,\psi_V(p) \,, \\
&& \int|\psi_V(p)|^2\,\dfr^3 p=1 \,, \qquad
\int|\psi_V(\beta,p_T)|^2\dfr^2 p_T\dfr \beta  = 1 \,,
\nonumber
\end{eqnarray}
in terms of the heavy-quark 3-momentum $p\equiv |\vec p\,|$. This prescription, also known 
as the Terent'ev recipe, has been found to be quite successful in describing the HERA 
data on exclusive electro- and photoproduction of charmonia states in
Refs.~\cite{Hufner:2000jb,Krelina:2018hmt,Cepila:2019skb}.
A justification of this prescription has been discussed in Ref.~\cite{Kopeliovich:2015qna} 
and no significant deviation in predictions has been found between the exact calculation and 
the Terent'ev recipe in the phenomenologically relevant domains of the phase space.

The spatial momentum-space wave function $\psi_V(p)$ is found by a Fourier-transform from the 
coordinate-space radial wave function $\psi(R)$ where $R \equiv |\vec{R}\,|$ is the interquark 
separation. The latter is found as a solution of the Schroedinger equation for a given interquark 
$Q-\bar Q$ interaction potential. In numerical calculations we employ five distinct models for 
the interquark potential: harmonic oscillator (osc), power-like model \cite{Martin:1980jx,Barik:1980ai} (pow), Buchmuller-Tye parametrisation \cite{Buchmuller:1980su} (but), Cornell potential \cite{Eichten:1978tg,Eichten:1979ms} (cor) and logarithmic potential \cite{Quigg:1977dd} (log). In each such parametrisation of the long-distance interaction potential, the heavy quark mass $m_Q$ ($Q=c,b$) was considered as an adjustable parameter fitted to describe the charmonia and bottomonia spectra. In the short-distance production amplitude, however, we use in numerical analysis the universal values for the charm ($m_c=1.4$ GeV) and bottom ($m_b=4.75$ GeV) 
quark masses.

\subsection{Dipole formula for photoproduction amplitude}
\label{sec:dipole}

Using the LF quarkonia wave function in Eq.~(\ref{eq:PsiV-LF}), the resulting 
photoproduction amplitude (\ref{psi-amp}) can be represented in 
the following form~\cite{Cepila:2019skb}
\begin{eqnarray}
&& \mathrm{Im}\mathcal{A}^{\gamma p\rightarrow V p}(W) = 
\int\limits_0^1\dfr \beta \int\dfr^2r_\perp \left[\Sigma^{(1)}(\beta, r_\perp)
\sigma_{q\bar q}(x,r_\perp) +
\Sigma^{(2)}(\beta , r_\perp)\frac{\dfr \sigma_{q\bar q}(x,r_\perp)}{\dfr r_\perp}\right]\,,
\label{ImA}
\end{eqnarray}
where the coefficient functions read
\begin{eqnarray}
&& \Sigma^{(1)} = \frac{Z_Q\sqrt{N_c\alpha_{\rm em}}}{2\pi\sqrt{2}}\,
2 K_0(m_Q r_\perp)\int \dfr p_T J_0(p_Tr_\perp)\psi_{V}(\beta ,p_T)
p_T\,\frac{m_Tm_L+m_T^2-2\beta (1-\beta )p_T^2}{m_L+m_T} \,, \nonumber \\
\end{eqnarray}
and
\begin{eqnarray}
\Sigma^{(2)} = \frac{Z_Q\sqrt{N_c\alpha_{\rm em}}}{2\pi\sqrt{2}}\,
2K_0(m_Q r_\perp)\int \dfr p_T J_1(p_Tr_\perp)
\psi_{V}(\beta ,p_T)\frac{p_T^2}{2}\,\frac{m_L+m_T+(1-2\beta )^2m_T}{m_T(m_L+m_T)} \,. 
\nonumber
\end{eqnarray}
Here, $\alpha_{\rm em}$ is the fine structure constant, $N_c=3$ is the number 
of colors in QCD, $Z_Q$ is the electric 
charge of the heavy quark, $J_{0,1}$ ($K_0$) are 
the (modified) Bessel functions of the first (second) kind, respectively, 
$p_T$ is the transverse momentum of the produced quarkonium state, and
\begin{eqnarray}
m_T = \sqrt{m_Q^2 + p_T^2} \,, \quad m_L = 2m_Q\,\sqrt{\beta (1-\beta )} \,.
\end{eqnarray}

Finally, in order to take into account the corrections due to the real part of the forward photoproduction amplitude we included an extra factor which is the ratio of the real to imaginary parts of the scattering amplitude \cite{Hufner:2000jb}
\begin{equation}
    \mathcal{A}^{\gamma p\rightarrow V p}(W) =   \mathrm{Im} 
    \mathcal{A}^{\gamma p\rightarrow V p}(W) \left(1 - i \frac{\pi}{2} 
    \frac{\partial \ln \mathrm{Im} \mathcal{A}^{\gamma p\rightarrow V p}(W) }
    {\partial \ln W^2} \right) \,.
\end{equation}
However, since the derivative of the imaginary part of the amplitude is sensitive only to $x$, we found it more convenient (see Refs.~\cite{Ivanov:2002kc}) to rewrite this expression in terms of the dipole cross section 
\begin{equation}
         \sigma_{q\bar{q}}(x,r_\perp) \Rightarrow \sigma_{q\bar{q}}(x,r_\perp) \left(1 - i \frac{\pi}{2}\frac{\partial \ln \sigma_{q\bar{q}}(x,r_\perp) }
     {\partial \ln W^2}  \right) \, ,
\end{equation}
that will be considered in further calculations.

\subsection{Saturated dipole cross section}
\label{sec:sigma_qq}

An essential ingredient of the photoproduction amplitude (\ref{ImA}) at high energies 
is the universal dipole cross section $\sigma_{q\bar q}(x,r_\perp)$ related to the gluon 
distribution in the proton target. At very low-$x$, one expects to enter 
a non-linear QCD evolution regime known as saturation that constrains the maximum 
of the gluon density that can be achieved in the hadronic wavefunction (see e.g. Refs.~\cite{Weigert:2005us,JalilianMarian:2005jf,Gelis:2010nm} and references therein). 
This effect is typically accounted for in phenomenological parametrisations for 
the dipole cross section whose saturated shape is characterised by the $x$-dependent 
saturation scale, $Q_s(x)$. 

One of such simplest and most phenomenologically successful saturated models is known 
as the Golec-Biernat--Wustoff (GBW) \cite{GolecBiernat:1998js} parametrisation,
\begin{eqnarray}
\label{GBW-model}
\sigma_{q{\bar q}}(x,r_\perp) = \sigma_0 \,
\Big(1 - e^{-\frac{r_\perp^2 Q_s^2(x)}{4}}\Big) \,,
\end{eqnarray}
satisfying the renown color transparency property of the dipole scattering, 
$\sigma_{q\bar q}(r_\perp) \propto r_\perp^2$ as $r_\perp\to 0$. Besides, we stick to the standard assumption
about the quark flavor invariance of the dipole cross section in this study. In Eq.~(\ref{GBW-model}), the parametrisation of the saturation scale squared in the proton target case
\begin{eqnarray}
Q_s^2(x) \equiv R_0^{-2}(x) = Q_0^2\Big( \frac{x_0}{x} \Big)^\lambda \,,
\label{proton-saturation}
\end{eqnarray}
is valid at very small $x\lesssim 0.01$ only. The saturated ansatz (\ref{GBW-model}) has
given rise to a whole family of dipole models attempting in particular to incorporate
the hard scale dependence of the dipole cross section via e.g. QCD DGLAP-like evolution. 
In the case of photoproduction, however, an effect of such a scale dependence remains
minor for not very large masses of the produced states and will be safely ignored 
in what follows. An early fit of the HERA data for the GBW model parameters in
Eq.~(\ref{GBW-model})
\begin{eqnarray}
\label{GBW-par}
Q_0^2 = 1\, {\rm GeV}^2\,, \quad x_0 = 3.04 \times 10^{-4}\,, \quad 
\lambda = 0.288\,, \quad \sigma_0 = 23.03\, {\rm mb} \,.
\end{eqnarray}
has been performed in Ref.~\cite{GolecBiernat:1999qd} yielding a good description 
of a big variety of various observables in both $ep$ and $pp$ collisions at high 
energies. 

Motivated by an analysis of theoretical uncertainties performed in
Ref.~\cite{Cepila:2019skb}, in what follows we use another parametrisation 
from Ref.~\cite{Kopeliovich:2003cn} 
known as the KST model providing a reasonably good description of the real 
photoproduction data available from the HERA collider. When $Q^2\to 0$ the Bjorken variable 
$x$ becomes inappropriate, such that $\sigma_0$ and $R_0$ in Eq.~(\ref{GBW-model}) 
should be replaced by functions $\bar{\sigma}_0(\hat s)$ and $\bar{R}_0(\hat s)$ 
of the dipole-target collision center-of-frame energy squared $\hat{s}\equiv W^2$ 
found as follows (for more details, see Ref.~\cite{Kopeliovich:1999am})
\begin{eqnarray}
 \bar{R}_0(\hat s)=0.88\,{\rm fm}\,(s_0/\hat s)^{0.14}\,,
\qquad
 \bar{\sigma}_0(\hat s)=\sigma_{\rm tot}^{\pi p}(\hat s)
 \Big(1+\frac{3\bar{R}_0^2(\hat s)}{8\langle r_{\rm ch}^2 \rangle_{\pi}}\Big)\,,
 \label{KST-params}
\end{eqnarray}
respectively, where
\begin{eqnarray}
\sigma_{\rm tot}^{\pi p}(\hat s)=23.6(\hat s/s_0)^{0.08}\,{\rm mb} \,, \qquad
\langle r_{\rm ch}^2 \rangle_{\pi}= 0.44\,{\rm fm}^2 \,,
\end{eqnarray}
are the total pion-proton scattering cross section and the mean pion radius squared 
\cite{Amendolia:1984nz}, respectively. Here, $s_0=1000\,{\rm GeV}^2$. The KST model
is generally considered to be applicable for soft and semi-hard processes up to 
scales of $Q^2\sim 20$ GeV$^2$ or so and, thus, particularly suitable for the 
predominantly soft charmonia photoproduction observables.

The effects of the skewness in the unintegrated gluon density when the gluons attached
to a quark-antiquark pair carry different light-front momentum fractions $x'\ll x \ll 1$ 
of the proton momentum are typically accounted for by an overall multiplicative correction 
factor slowly dependent on gluon kinematics \cite{Shuvaev:1999ce, Martin:1999wb}.
The status and the exact analytic form of the skewness correction in the dipole picture
of the elastic quarkonia photoproduction are not fully understood in the literature
within the kinematic ranges studied in the current analysis.

In our analysis of quarkonia photoproduction in UPCs, following 
Refs.~\cite{Ivanov:2002kc,Hufner:2000jb} we employ only GBW and KST 
models described above. This is also motivated by an observation of Ref.~\cite{Cepila:2019skb} 
that these two models provide similar results at low-$x$ and both lead to a reasonably 
good description of the charmonia photoproduction data at center-of-mass energies 
$W\lesssim 200\,\GeV$ available from the HERA collider without any additional factors. 

In fact, a detailed discussion of various theoretical uncertainties in the considered 
processes has been given in Ref.~\cite{Cepila:2019skb}, including an analysis of 
the skewness correction, and we follow the same reasoning in the current work. 
It was shown there, in particular, that the use of skewness factor typically increases 
the photo- and electroproduction cross section of quarkonia by a factor of $1.5 - 1.6$. 
Besides, it was demonstrated that, omitting the skewness correction, only the KST and GBW 
dipole parametrizations work well against the HERA data on exclusive quarkonium 
electroproduction, while all other known phenomenological dipole cross sections 
noticeably underestimate these data. An effort to obtain a better agreement with the data 
in these models typically provide the main reason to include formally the skewness 
effects adopting only an approximate factorised expression for the skewness 
factor \cite{Shuvaev:1999ce}. 
However, this is based on assumptions which may not be 
naturally adopted or justified for an arbitrary process and we avoid making such 
assumptions in the current work. In turn, the successful use of the KST 
\cite{Kopeliovich:2003cn} and GBW \cite{GolecBiernat:1998js} dipole parametrizations 
off the proton target motivates us to use the same approach also for nuclear targets in UPCs.

Strictly speaking, the dipole parameterisations discussed above contain only the part 
of the gluon density that increases at low-$x$. At large $x>0.01$, however, the gluon density in the target decreases approximately as $g(x)\propto (1-x)^N$ suggested by the dimensional-cutting rules \cite{Drell:1969km,West:1970av,Brodsky:1973kr}, where $N\sim 5\div 8$ depending on the hard scale of the process. A multiplication of the saturation scale squared $Q_s^2(x)$ by such a kinematical threshold factor $(1-x)^N$ is often referred to as the modified dipole approach that is known to provide a significant improvement of the Drell-Yan data description at large $x$ (while the small-$x$ regime is practically unaffected) \cite{Kopeliovich:2002yh,Raufeisen:2002zp} (see also Ref.~\cite{Kutak:2003bd}). Along these lines, in our numerical analysis we supplement the dipole cross section with a factor $(1-x)^{2n_s - 1}$, where $n_s$ is the number of the active spectator quarks for the process (we adopt $n_s = 4$ in this work).

\subsection{Numerical results for $\gamma p\to V p$ cross sections}
\label{sec:results-gammap}

\begin{figure*}[!htbt]
\begin{minipage}{0.48\textwidth}
 \centerline{\includegraphics[width=1.0\textwidth]{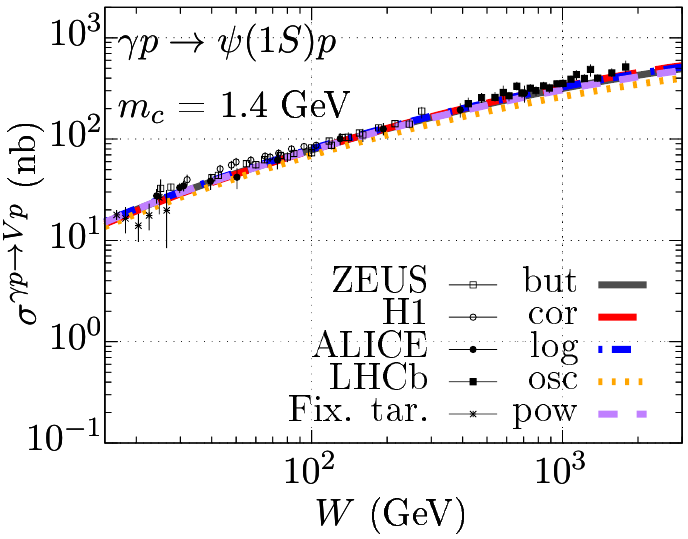}}
\end{minipage} \hfill
\begin{minipage}{0.48\textwidth}
 \centerline{\includegraphics[width=1.0\textwidth]{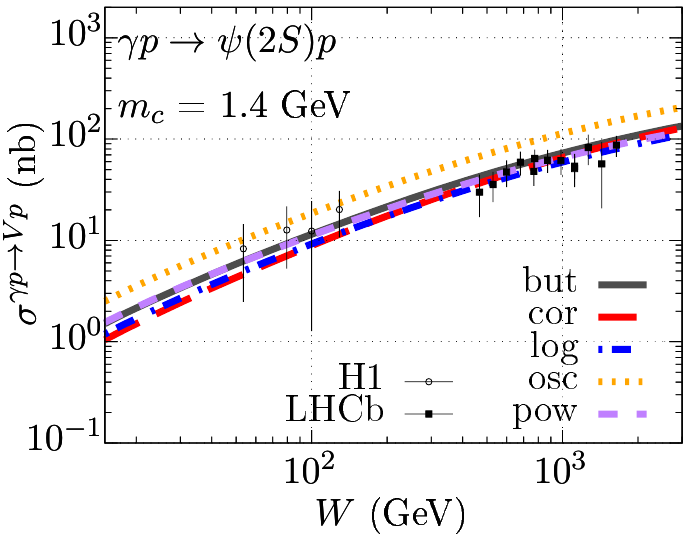}}
\end{minipage} \hfill
\caption{Integrated diffractive $\gamma p\to V p$ photoproduction cross section as a function 
of $\gamma p$ center-of-mass energy, $W$, for $V=\psi(1S)$ (left) and $V=\psi(2S)$ (right) 
using the GBW dipole parametrisation (\ref{GBW-par}). The results are compared with 
the available experimental data from H1 \cite{Alexa:2013xxa}, ZEUS \cite{Chekanov:2002xi}, 
ALICE \cite{Abelev:2012ba} and LHCb \cite{Aaij:2018arx} collaborations as well as from 
the fixed-target experiment at Fermilab \cite{Frabetti:1993ux,Denby:1983az,Binkley:1981kv}.
}
\label{fig:sigma-psix_W}
\end{figure*}

Let us now turn to a discussion of numerical results for the integrated 
diffractive $\gamma p\to V p$ photoproduction cross sections (i.e. with the proton 
target), for $V=\psi(nS),\Y(nS)$, $n=1,2$. In Fig.~\ref{fig:sigma-psix_W}, we present
the dipole model results for $\psi(1S)$ (left panel) and $\psi(2S)$ (right panel) 
cross sections as functions of $\gamma p$ center-of-mass energy, $W$. In this analysis, 
we have used five different models for the interquark potential available 
from the literature and mentioned earlier. We notice that for charmonia photoproduction 
both dipole parametrisations, 
GBW and KST, discussed above in Sect.~\ref{sec:sigma_qq} give very similar results so we have 
chosen the GBW parametrisation for the presentation purposes here. Our results are compared 
to the data available from H1 \cite{Alexa:2013xxa}, ZEUS \cite{Chekanov:2002xi}, ALICE
\cite{Abelev:2012ba} and LHCb \cite{Aaij:2018arx} measurements as well as from the 
fixed-target measurements at Fermilab \cite{Frabetti:1993ux,Denby:1983az,Binkley:1981kv}.
One observes that all five potentials used in our calculations give a relatively good 
description of the data for diffractive photoproduction of both $\psi(1S)$ and 
$\psi(2S)$ states in the considered energy range.
\begin{figure*}[!htbt]
\begin{minipage}{0.48\textwidth}
 \centerline{\includegraphics[width=1.0\textwidth]{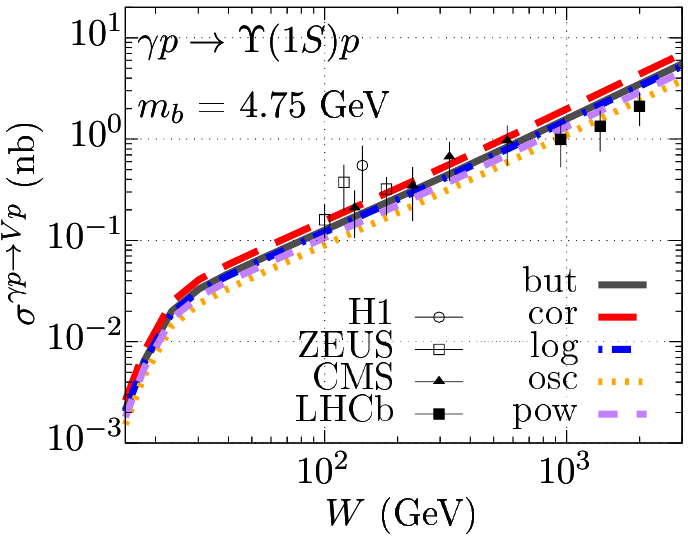}}
\end{minipage} \hfill
\begin{minipage}{0.48\textwidth}
 \centerline{\includegraphics[width=1.0\textwidth]{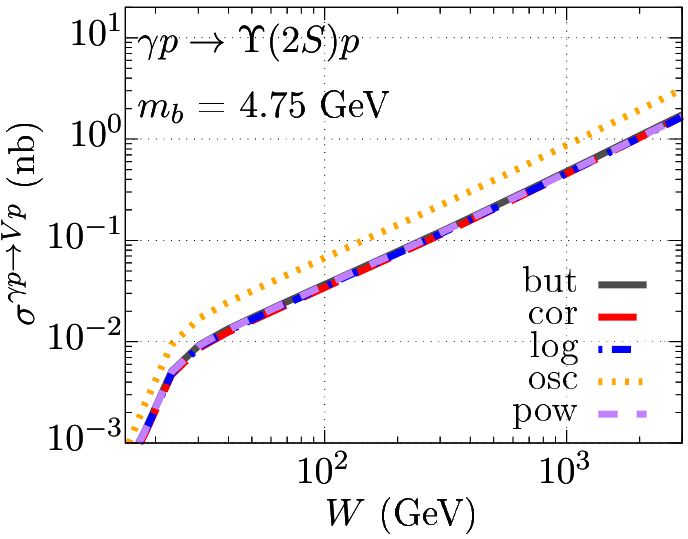}}
\end{minipage} \hfill
\caption{Integrated diffractive $\gamma p\to V p$ photoproduction cross section as a function 
of $\gamma p$ center-of-mass energy, $W$, for $V=\Y(1S)$ (left) and $V=\Y(2S)$ (right) 
using the KST dipole parametrisation (\ref{GBW-par}). The results are compared with 
the available experimental data from CMS \cite{Sirunyan:2018sav}, H1 \cite{Adloff:2000vm}, 
ZEUS \cite{Breitweg:1998ki,Chekanov:2009zz} and LHCb \cite{Aaij:2015kea} collaborations.}
\label{fig:sigma-upsilonx_W}
\end{figure*}

In a separate dedicated analysis, we have compared the numerical results for the integrated 
diffractive $\gamma p\to V p$ photoproduction cross section obtained with the original GBW
model \cite{GolecBiernat:1998js} discussed above and with the updated GBW fit accounting for heavy quarks \cite{Golec-Biernat:2017lfv}. We notice that the numerical difference between the results obtained 
with these two sets of GBW parametrizations is generally very small. Another observation is that 
for large values of $W$, the results obtained with the original GBW parameterisation 
are somewhat closer to the data points. This is the reason why we have chosen
``old'' GBW fit from Ref.~\cite{GolecBiernat:1998js} in our current analysis.

In Fig.~\ref{fig:sigma-upsilonx_W}, we show the numerical results for the integrated 
cross sections of diffractive $\Y(1S)$ (left panel) and $\Y(2S)$ (right panel) 
photoproduction as functions of $W$. In analogy to the previous figure, in our 
calculations of the radial wavefunction of the bottomonia states, we employed
five different models for the $b\bar b$ interaction potential. The results for 
the ground state are confronted against the available $\Y(1S)$ photoproduction data 
from CMS \cite{Sirunyan:2018sav}, H1 \cite{Adloff:2000vm}, ZEUS 
\cite{Breitweg:1998ki,Chekanov:2009zz} and LHCb \cite{Aaij:2015kea} collaborations.
In this figure, we have shown the results with the KST parametrisation of the dipole 
cross section (\ref{KST-params}) since it provides the best description of $\Y$ 
photoproduction data. It is worth noticing that all five potentials provide 
a comparatively good description of the available data on $\Y(1S)$ in the considered 
energy range. This is the reason why we have used the same dipole model parametrisation 
and the interquark potentials for making predictions for the photoproduction cross 
section of the excited $\Y(2S)$ state shown in the right panel. For reliable and thorough
estimates of underlying theoretical uncertainties in our calculations, we refer the reader
to Ref.~\cite{Cepila:2019skb} where such uncertainties have been discussed in detail.

A close inspection of Fig.~\ref{fig:sigma-upsilonx_W} reveals that the difference between 
the $\gamma p $ production cross section for the ground and excited states decreases 
with the rise of $W$ for all potentials. In the case of oscillator potential, 
for example, the result for the $\Upsilon(1S)$ is approximately $22\%$ higher 
than the one for $\Upsilon(2S)$ at $W=1000$ GeV, being the smallest difference compared
to the other potentials. The observation that the oscillator potential gives relatively 
similar results for the ground and excited quarkonia is due to the a very similar small-$r$ 
dependence and magnitude of the corresponding light-front quarkonia wave functions computed 
with the oscillator potential (see Ref.~\cite{Cepila:2019skb} for more quantitative 
details). Such a dependence of the oscillator wave function is rather different 
compared to the other models. Since the main contribution to the integrated cross 
section comes from the region of small $r$ in the radial wave function, this effect 
is particularly relevant for $\Upsilon$ states and less so for charmonia.

\section{Exclusive photoproduction off the nucleus target}
\label{sec:QQ-off-nucleus}

Let us now turn to analysis of exclusive $\psi(nS),\Y(nS)$, $n=1,2$ quarkonia 
photoproduction off the heavy nucleus target $A$ relevant for the corresponding 
recent measurements in $AA$ UPCs at the LHC. Here, we briefly overview an extension of 
the dipole model framework and the formalism of LF quarkonia wavefunctions to this case,
as well as study the corresponding observables and confront them with all currently 
available data.

\subsection{Differential $AA \rightarrow AVX$ cross section and the photon flux}
\label{sec:photon-flux}

In exclusive quarkonia photoproduction off the proton target considered above, the experimental 
data are typically provided for $\sigma_{\gamma p\rightarrow Vp}$, since the photon flux, in this
case, is factorised from the elastic $\gamma p\rightarrow Vp$ cross section. However, such
factorisation does not hold in the case of a nucleus target. Indeed, a nontrivial impact 
parameter dependence of the photon flux becomes relevant in the $AA$ UPCs. In the considering 
kinematics, the photons can only interact with the target when there is no overlap between 
the projectile and the target in impact parameter $b$-space such that $b > 2 R_A$, in terms 
of the nucleus radius $R_A$ (for more details and review, see Ref.~\cite{Baltz:2007kq} and 
references therein). We also define $b'$ as the position of the interaction point with 
respect to the target nucleus center, and $\vec{b}_\gamma = \vec{b}' - \vec{b}$ is the same 
position but with respect to the projectile nucleus. 

In the center-of-mass frame of the colliding particles, the produced quarkonium state $V$ 
has mass $M_V$ and rapidity $y$. Also, the photon energy is labelled as $\omega$ and 
\begin{equation}
    \omega \approx \frac{M_V}{2}e^y = \frac{W^2}{2\sqrt{s}} \,,
    \label{photon_energy}
\end{equation}
where $\sqrt{s}$ is the projectile-target (with the projectile being a nucleon inside of 
an incoming nucleus and the target -- a nucleon inside the target nucleus) center-of-mass 
energy. Of course, in numerical computations of the differential cross section one 
has to take into account that the photon can be originated from each of the nuclei, 
which is done by incorporating $y \rightarrow -y$. 

The differential cross section of quarkonia photoproduction in $AA$ UPCs when the
projectile photon is taken from one of the colliding nuclei is given in the following 
standard form \cite{Ivanov:2002eq}
\begin{equation}
\frac{d\sigma^{AA \rightarrow AVX}}{dy} 
 = \int \dfr^2b \int \dfr^2b' \,
    \omega\frac{d N_\gamma(\omega, \vec{b}_\gamma)}{d\omega d^2 b_\gamma} \, 
    \frac{d\sigma^{\gamma A \rightarrow VX}(\omega,\vec b')}{d^2b'} \,, \qquad
    \vec{b}_\gamma = \vec{b}' - \vec{b} \,,
    \label{dif_cross_rap}
\end{equation}
where $X=A$ or $A^*$ for the coherent and incoherent production, respectively. 
Provided that the photons are quasi-real in the considering UPCs, i.e.
$Q^2\approx 0$, only transversely polarised photons are relevant. The photon 
number density in the projectile nucleus is conventionally described by 
the differential Weisz\"acker-Williams (WW) photon flux
\cite{vonWeizsacker:1934nji,Williams:1934ad}, with a broad spectrum
as given by
\begin{eqnarray}
\frac{d^3N_\gamma(\omega,\vec{b}_\gamma)}{d\omega d^2b_\gamma} = 
\frac{Z^2\alpha_{\rm em} k^2}{\pi^2\omega b_\gamma^2} 
\Big[ K^2_1(k)+\frac{1}{\gamma^2}K^2_0(k) \Big]\,, 
\qquad k=\frac{b_\gamma\, \omega}{\gamma} \,,
\label{WW-flux}
\end{eqnarray}
where $\gamma = \sqrt{s}/2m_p$ is the Lorentz factor, the proton mass is given 
by $m_p=0.938$ GeV, and $Z$ is the charge of the projectile nucleus. Since at the LHC 
energies the Lorentz factor is large $\gamma\gg 1$ (e.g. in 2016 $pPb$ run with 
$\sqrt{s} = 8.16$\,TeV, $\gamma_{\rm Pb} \approx 4350$), the second term 
in Eq.~(\ref{WW-flux}) can be safely omitted in practical calculations.

\subsection{Coherent and incoherent processes}
\label{sec:coh-inc}

In the $\gamma A \rightarrow V X$ subprocess cross section entering in Eq.~(\ref{dif_cross_rap}) 
one separates two distinct vector meson production modes. The first one, called coherent 
production, occurs when the nucleus target remains intact after the interaction, i.e. $X=A$. 
In this case, the production cross section is computed in the framework of the Good-Walker 
formalism \cite{Good:1960ba} by averaging the color dipole interactions over all 
possible configurations of the projectile nucleus, thus, probing the average distribution 
of low-$x$ gluons in the target \cite{Miettinen:1978jb,Kowalski:2006hc}.
The second process called incoherent production occurs when the outgoing target nucleus does 
not retain the same quantum state as the incoming nucleus, i.e.~it becomes an excited state $A^*$, 
which contains nucleons and nuclear fragments but no other hadrons. This process features
a large gap in rapidity between the produced quarkonium and $A^*$ system, and measures how much 
the scattering amplitude fluctuates between the different possible initial-state configurations
\cite{Caldwell:2009ke} (for a recent detailed discussion, see e.g.~Ref.~\cite{Mantysaari:2020axf}).

A number of different approaches for a detailed treatment of the incoherent photo-nuclear 
production have been developed in the literature so far, see e.g. Refs.~\cite{Klein:2016yzr,Luszczak:2019vdc,Luszczak:2017dwf,Chen:2018vdw,Goncalves:2017wgg,Sambasivam:2019gdd,Mantysaari:2017dwh,Cepila:2017nef,Guzey:2018tlk,Ducati:2017bzk,Ducati:2016jdg}. However, an adequate simultaneous description of both coherent and incoherent processes, 
in the framework of the same approach, remains an open problem in the literature.
Looking specifically into the incoherent case, Ref.~\cite{Mantysaari:2017dwh} provides a reasonable description of the only available data point for $J/\psi$ photoproduction in PbPb UPCs at 2.76 TeV from ALICE Collaboration \cite{Abbas:2013oua}, but when the same approach is used for treatment of the coherent case, it does not describe the data so well. In our work, we are primarily concerned about getting an accurate description of the coherent case (in particular, coherent $J/\psi$ photoproduction at $\sqrt{s} = 2.76$ TeV), and then we employ the same phenomenological framework for treatment of the incoherent production as well.

Compared to the proton target case considered above, two additional effects are known to play
a critical role in quarkonia photoproduction off a heavy nucleus \cite{Kopeliovich:1991pu,Ivanov:2002eq,Ivanov:2002kc}. 
The first one is called the color filtering i.e. inelastic scatterings of the $Q\bar Q$ pair 
in the course of its propagation through the nucleus. The second effect is associated with 
the nuclear shadowing of the gluon density due to a reduction of the dipole scattering 
cross section off the nucleus target compared to that off the proton due to interferences.
Both effects cause a reduction of quarkonia photoproduction off the nuclear target, 
$\gamma A \rightarrow V X$, compared to $A\sigma^{\gamma p \rightarrow V p}$, and are effectively accounted for in our analysis below.

At high energies, the coherence (or production) length $l_c$ defined as 
the lifetime of the $q \bar q$ fluctuation \cite{Kopeliovich:1991pu,Brodsky:1988xz}, 
\begin{equation}
l_c = \frac{2\omega'}{M_V^2} \,,
\end{equation}
is often considered to be much larger than the nuclear radius, $l_c \gg R_A$, where $\omega'$ is the photon energy in the target rest frame. 
This guaranties a small variation of the transverse size of the dipole system 
and no fluctuations during the propagation process through the nucleus by means 
of Lorentz time dilation. In this case, a $Q\bar Q$ Fock fluctuation of the photon 
builds up long before it interacts with the nucleus target. This is called 
the ``frozen'' approximation, and the calculations are particularly simple 
and well-known in this case. 

Note, there is yet another scale called the formation length of the heavy quarkonia defined 
as $l_f=2\omega'/(M_{2S}^2-M_{1S}^2)$ \cite{Kopeliovich:1991pu,Brodsky:1988xz}, 
which is larger than the coherence length scale. In the considered case of 
exclusive photoproduction, the uncertainty principle enables to resolve between $J/\psi$
and $\psi'$ as long as the formation length $l_f$ is smaller than the mean inter-nucleon 
separation in the target nucleus. Hence, at high energies, when $l_f\gtrsim R_A$, the use of 
a different approach than a simple eikonalization of the photoproduction off the nucleon 
target is necessary.

In the ``frozen'' $l_c\to \infty$ limit, the incoherent and coherent production cross 
sections are found as \cite{Kopeliovich:1991pu,Ivanov:2002eq,Ivanov:2002kc}
\begin{eqnarray}
\sigma^{\gamma A \rightarrow V A^*} &=& \int \dfr^2 b 
\frac{T_A(b)}{16 \pi B}\left| 
\int \dfr\beta \dfr^2 r_\perp  \Psi_{V}^{\dagger} 
\Psi_\gamma \sigma_{q\bar q}(x,r_\perp) 
\exp\left(-\frac{1}{2}\sigma_{q\bar q}(x,r_\perp)T_A(b)\right)\right|^2 \,,
\label{CS-incoh-large-lc} \\
\sigma^{\gamma A \rightarrow V A} &=& \int \dfr^2 b \left| 
\int \dfr\beta  \dfr^2 r_\perp
\Psi_{V}^{\dagger} \Psi_\gamma \left[1 - \exp\left(-\frac{1}{2}
\sigma_{q\bar q}(x,r_\perp) 
T_A(b) \right) \right]\right|^2 \,, \label{CS-coh-large-lc}
\end{eqnarray}
respectively, where $\Psi_{V}=\Psi_{V}(\beta,r_\perp)$ and 
$\Psi_{\gamma}=\Psi_{\gamma}(\beta,r_\perp)$ are the LF vector meson 
and real photon wave functions discussed above, and $T(b)$ 
is the thickness function of the nucleus
\begin{eqnarray}
T(b) = \int_{-\infty}^{+\infty} \dfr z \, 
\rho_A(b,z) \,, \qquad \int \dfr^2 b \, T(b) = 1 \,,
\end{eqnarray}
defined as an integral of the LC nuclear density, $\rho_A(b,z)$, over the longitudinal
coordinate $z$ at a fixed impact parameter $b$. In this paper, for the latter we employ 
the Woods-Saxon parametrisation \cite{Woods:1954zz}
\begin{eqnarray}
\rho_A(b,z) = \frac{N_A}{1 + \exp[\frac{r(b,z) - R_A}{\delta}]} \,, \qquad
r(b,z)=\sqrt{b^2 + z^2} \,,
\end{eqnarray}
where $r\equiv |\vec r|$ is the distance from the center of the nucleus,
$N_A$ is an overall normalization factor, and the parameters $R_A = 6.62$ fm and 
$\delta = 0.546$ fm are taken from Ref.~\cite{Euteneuer:1978qw}. 

\subsection{Finite coherence length}
\label{sec:fin-coh-length}

Effects due to a finite coherence length, i.e. when $l_c < R_A$, become particularly 
relevant at low energies $W$. In this case, the real photon propagates inside the nucleus 
target not experiencing any attenuation until it develops a $Q\bar Q$ fluctuation (at a short 
time scale $t_c$) which then instantly interacts with the nuclear medium. In the analysis 
of these dynamics, one should take into account that the produced $Q\bar Q$ dipole attenuates 
along its propagation path in the nuclear absorptive medium, whose typical mean length 
is roughly a half of the nuclear thickness \cite{Ivanov:2002kc}.

A manifest quantum-mechanical mechanism for the propagation of such a fluctuating $Q\bar Q$ dipole through
the nuclear medium is consistently implemented in the framework of the LC Green function approach 
\cite{Kopeliovich:2001xj}. An analytical solution for dipole Green function is known only for the
simplest quadratic dependence of the dipole cross section, $\sigma_{q\bar q}\propto r^2$, 
while a generic case is treatable only numerically and is rather challenging.

Here we follow the approximation of Ref.~\cite{Hufner:1996jw} (with the explicit formulae given 
in Ref.~\cite{Ivanov:2002kc}) in which the effect of the finite coherence length is effectively 
taken into account by multiplying the infinite coherence length results (\ref{CS-incoh-large-lc}) 
and (\ref{CS-coh-large-lc}) by a corresponding form factor as
\begin{eqnarray}
    \sigma^{\gamma A \rightarrow V A^*} (W^2) \Rightarrow 
    \sigma^{\gamma A \rightarrow V A^*} (W^2) F^{\rm inc} (W^2, l_c) \,, \label{CS-incoh-small-lc} \\
    \sigma^{\gamma A \rightarrow V A} (W^2) \Rightarrow 
    \sigma^{\gamma A \rightarrow V A} (W^2) F^{\rm coh} (W^2, l_c) \,, \label{CS-coh-small-lc}
\end{eqnarray}
for the incoherent and coherent production cross sections, respectively. In the first, 
incoherent case, the form factor is represented in the form explicitly normalized to 
the $l_c = \infty$ case as follows
\begin{equation}
    F^{\rm inc} (W^2, l_c) = \int \dfr^2b \int^{\infty}_{-\infty} \dfr z 
    \rho_A(b, z) \Big|F_1(W^2, b, z) - F_2(W^2, b, z, l_c)\Big|^2 
    \Big/ \Big(...\Big)\Big|_{l_c \to \infty} \,. \label{FF-inc}
\end{equation}
Here, 
\begin{equation}
    F_1(W^2,b, z) = \exp \left( - \frac{1}{2} \sigma_{\rm VN}(W^2) \int_z^{\infty} \dfr z' 
    \rho_A(b,z')\right)
\end{equation}
takes into account that the incident photon fluctuates into the $Q\bar Q$ pair and interacts 
with nucleus at some point with longitudinal coordinate $z$, which then propagates through 
the nucleus and forms a vector meson. The latter then leaves without experiencing 
inelastic interactions. The above relation is written in terms of the energy-dependent 
meson-nucleon total cross section, $\sigma_{\rm VN}(W^2)$, since the result is dominated 
by the $Q\bar Q$ dipole propagation with typical sizes relevant for $Q\bar Q$ projection 
to a particular meson state $V$. The second contribution in Eq.~(\ref{FF-inc}) reads
\begin{equation}
    F_2(W^2,b,z,l_c) = \frac{1}{2}\sigma_{\rm VN}(W^2)\int_{-\infty}^z \dfr z' 
    \rho_A(b,z') F_1(W^2,b,z') e^{i(z'-z)/l_c} \,.
\end{equation}
It accounts for a possibility that the photon first elastically produces 
a quarkonium state at a point $z'$, $\gamma A \to V A$, which then propagates 
through the nucleus target without interacting till another point $z>z'$,
where the last (quasi-elastic) scattering occurs giving rise to the final-state 
quarkonium, $V A \to V A^*$.

In the coherent case, the fact that the mesons produced at different longitudinal coordinates 
and impact parameters add up coherently simplifies the expression for the form factor compared 
to the previous case yielding
\begin{equation}
    F^{\rm coh}(W^2,l_c) = \int \dfr^2 b \left|\int^{\infty}_{-\infty} \dfr z \rho_A(b, z) 
    F_1(W^2,b, z) e^{i z/l_c} \right|^2 \Big/ \Big(...\Big)\Big|_{l_c \to \infty} \,.
\end{equation}
A more sophisticated Green functions analysis of the coherence length effects against the recent 
data on quarkonia photoproduction in UPCs will be done elsewhere.

\subsection{Gluon shadowing}
\label{sec:G-shadowing}

At small $x$, the gluon density of a nucleon inside the target nucleus is suppressed compared
to that of a free nucleon -- the phenomenon known as the gluon (or nuclear) shadowing. In the 
target rest frame, this is understood as a result of the interference among incoming dipoles 
due to the presence of the higher Fock states of the projectile photon. The leading-order 
gluon shadowing correction emerges via eikonalization of the next Fock component of 
the photon containing the $Q\bar Q$ dipole plus a gluon. This effect can be 
accounted for by renormalizing the dipole cross section as follows
\cite{Kopeliovich:1999am,Ivanov:2002eq}
\begin{equation}
    \sigma_{q\bar{q}}(r,x) \rightarrow  \sigma_{q\bar{q}}(r,x) R_g(x, \mu^2) \,,
\end{equation}
where the factor $R_g$ represents the ratio of the gluon density inside a nucleon 
in the nucleus compared to the one inside the free proton, i.e.
\begin{equation}
   R_g(x, \mu^2) =  \frac{x g_A(x,\mu^2)}{A x g_p(x,\mu^2)} \,.
\end{equation}
In this work, we compute this factor phenomenologically using the EPPS16 nuclear parton 
distributions fitted to the LHC data \cite{Eskola:2016oht}, with $\mu = M_V/2$ as 
the factorization scale. At this point, our calculations differ from most of 
previous analyses in the literature which instead have used theoretical models to predict 
the amount of the gluon shadowing.

\subsection{Numerical results for $AA \to AVX$ cross sections}
\label{sec:results-AA}

\begin{figure*}[!htbt]
\begin{minipage}{0.48\textwidth}
 \centerline{\includegraphics[width=1.0\textwidth]{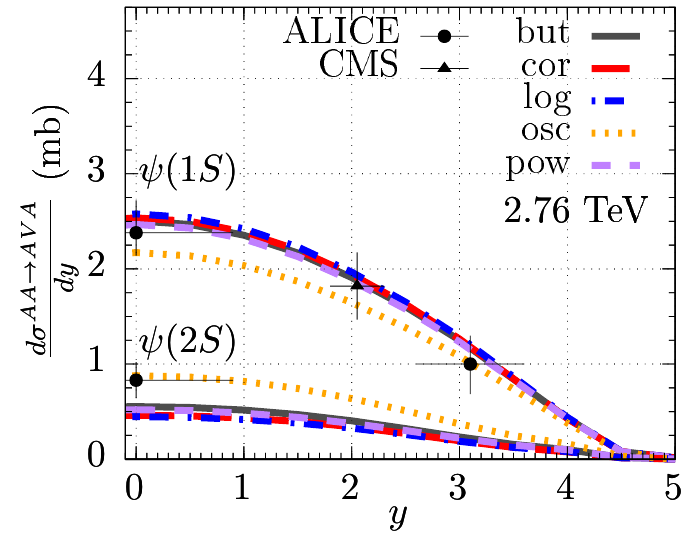}}
\end{minipage} \hfill
\begin{minipage}{0.48\textwidth}
 \centerline{\includegraphics[width=1.0\textwidth]{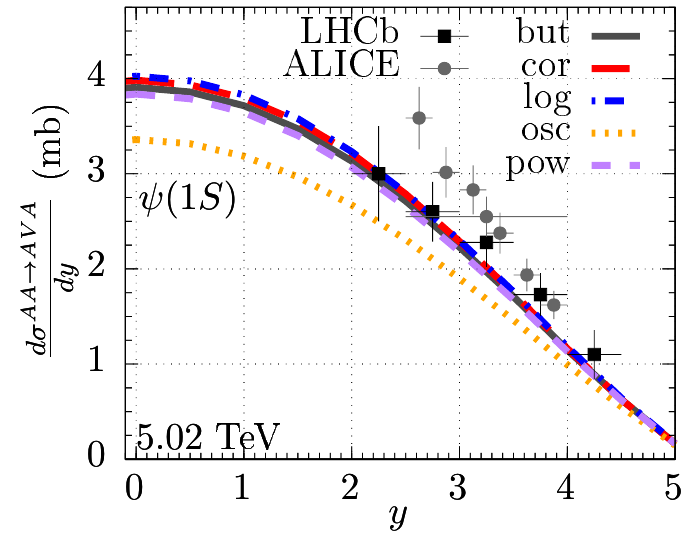}}
\end{minipage} \hfill
\caption{Differential (in rapidity) cross section for coherent $\psi$ photoproduction in PbPb UPCs 
at 2.76 TeV (left) and 5.02 TeV (right) using the GBW dipole model. 
The results with $2.76$ TeV are compared with data from CMS \cite{Khachatryan:2016qhq} 
and  ALICE \cite{Abbas:2013oua,Abelev:2012ba} for $J/\psi$ and from ALICE \cite{Adam:2015sia}
for $\psi'$, while the ones with $5.02$ TeV are compared with data from ALICE
\cite{Acharya:2019vlb} and preliminary results of LHCb \cite{LHCb:2018ofh}.}
\label{fig:sigmaAA-coh-psix-dy}
\end{figure*}

With all the formalism presented previously, we obtained some of the most relevant numerical 
results for the differential photoproduction cross sections in lead-lead collisions at the LHC energies. 
In Fig.~\ref{fig:sigmaAA-coh-psix-dy}, it is shown the differential cross sections for the coherent
photoproduction of $\psi(1S)$ and $\psi(2S)$ at $2.76$ TeV (left panel) and $\psi(1S)$ at $5.02$ TeV 
(right panel) as functions of the vector meson rapidity $y$. In these plots, we used again the five different models for the interquark
potentials and the GBW dipole model. In the left panel, we compared our results with the data from CMS
\cite{Khachatryan:2016qhq} and ALICE \cite{Abbas:2013oua,Abelev:2012ba} collaborations for $J/\psi$, 
and from ALICE collaboration \cite{Adam:2015sia} for $\psi'$. It can be noticed that our theoretical calculations describe very well the ground state production data with any of the considered interquark potentials.
However, in the case of the excited state production, only the result obtained with the oscillator $c\bar c$ potential reaches the corresponding data within error bars. In the right panel, the results are compared with the data from the ALICE collaboration
\cite{Acharya:2019vlb} and with the preliminary data from the LHCb collaboration \cite{LHCb:2018ofh}. 
In this case, our results provide a good description of the LHCb data, but fail to describe 
the ALICE data. We would like to point out that there is, in fact, a significant tension between 
these data sets themselves.
\begin{figure*}[!htbt]
\begin{minipage}{0.48\textwidth}
 \centerline{\includegraphics[width=1.0\textwidth]{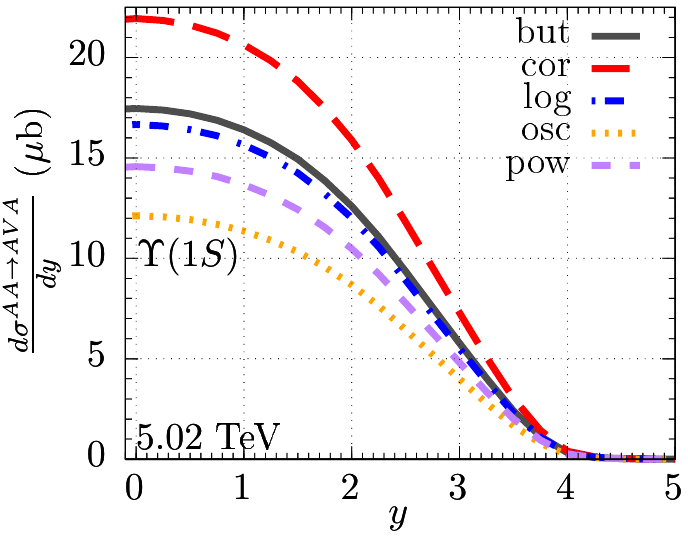}}
\end{minipage} \hfill
\begin{minipage}{0.48\textwidth}
 \centerline{\includegraphics[width=1.0\textwidth]{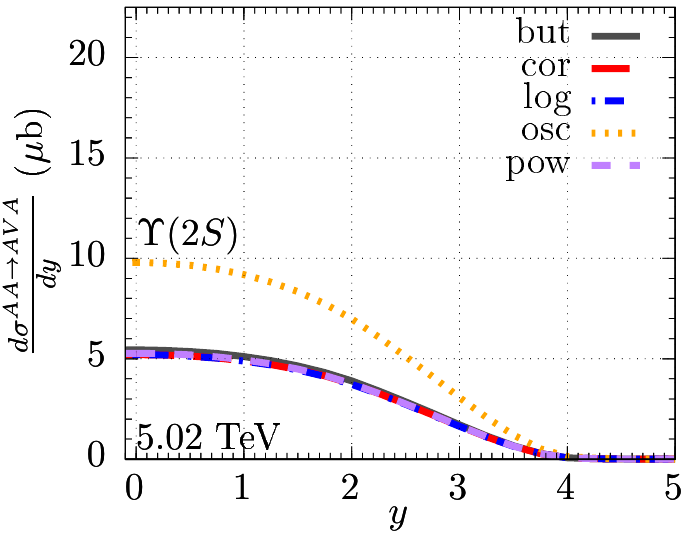}}
\end{minipage} \hfill
\caption{Differential (in rapidity) cross section for coherent $\Upsilon (1S)$ (left) 
and $\Upsilon (2S)$ (right) production at $5.02$ TeV in PbPb collision, using 
the KST dipole model.}
\label{fig:sigmaAA-coh-upsilonx-dy}
\end{figure*}

In Fig.~\ref{fig:sigmaAA-coh-upsilonx-dy}, we present our results for coherent 
photoproduction of $\Upsilon(1S)$ (left panel) and $\Upsilon(2S)$ (right panel) 
at $5.02$ TeV. The observable analyzed here is the differential cross section as a function 
of the meson rapidity calculated with the KST dipole cross section for the five distinct 
models of the $b\bar b$ potentials, in analogy to Fig.~\ref{fig:sigmaAA-coh-psix-dy}. 
As the reader can notice in the left panel, there is a significant spread between 
the results obtained with different potentials in the $\Upsilon(1S)$ production. 
This spread is especially pronounced at small values of the rapidity, since at large 
values of $y$ the effect of the finite coherence length dominates. This is a direct 
consequence of the shape of the LF $\Upsilon(1S)$ wave functions that differs more 
from one potential to another than the other vector meson wavefunctions. This does 
not occur in the production of the excited state, however, as can be seen in the right 
panel. In this case, there is a good agreement between the results obtained for each 
potential, except for the oscillator one which gives a higher prediction.
\begin{figure*}[!htbt]
\begin{minipage}{0.48\textwidth}
 \centerline{\includegraphics[width=1.0\textwidth]{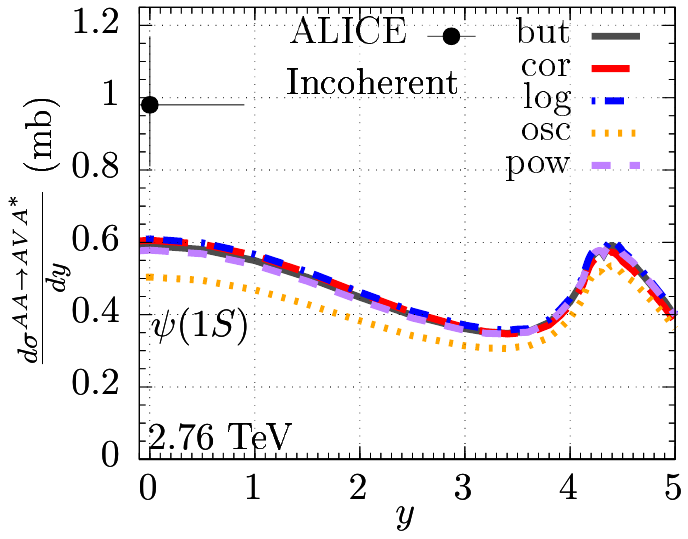}}
\end{minipage} \hfill
\begin{minipage}{0.48\textwidth}
 \centerline{\includegraphics[width=1.0\textwidth]{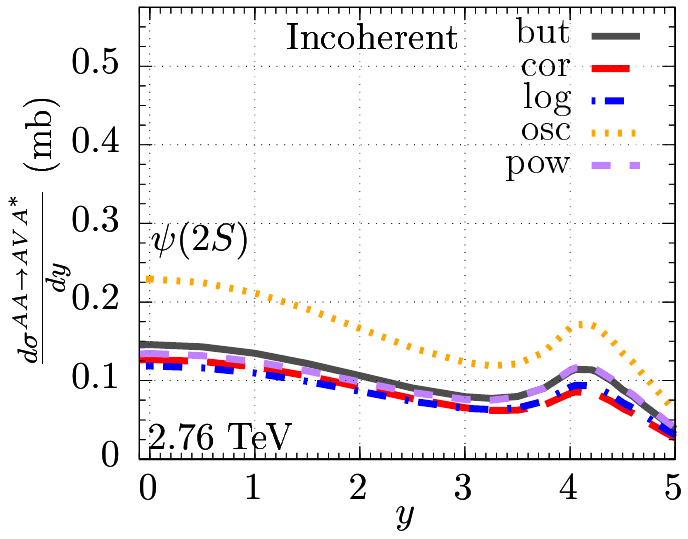}}
\end{minipage} \hfill
\caption{Differential (in rapidity) cross section for incoherent $\psi(1S)$ (left) 
and $\psi (2S)$ (right) photoproduction at $2.76$ TeV in PbPb UPCs. The ground state 
results are compared with the data from ALICE \cite{Abbas:2013oua}.}
\label{fig:sigmaAA-incoh-psix-dy}
\end{figure*}

\begin{figure*}[!htbt]
\begin{minipage}{0.48\textwidth}
 \centerline{\includegraphics[width=1.0\textwidth]{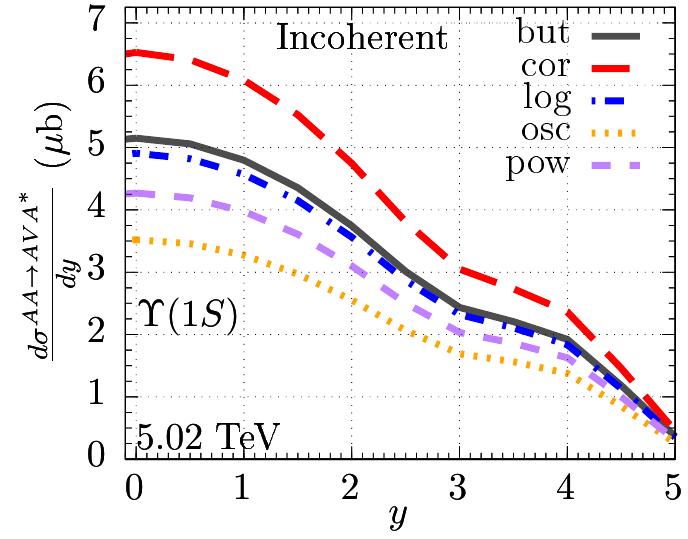}}
\end{minipage} \hfill
\begin{minipage}{0.48\textwidth}
 \centerline{\includegraphics[width=1.0\textwidth]{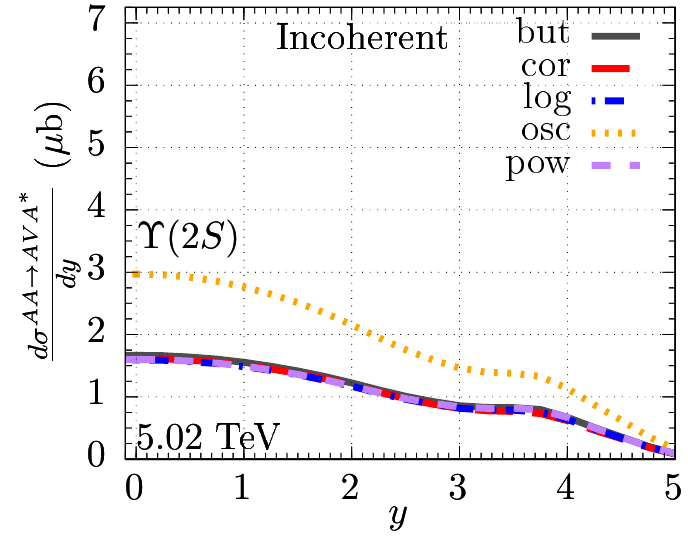}}
\end{minipage} \hfill
\caption{Differential (in rapidity) cross section for incoherent $\Upsilon (1S)$ (left) 
and $\Upsilon (2S)$ (right) photoproduction at $5.02$ TeV in PbPb UPCs.}
\label{fig:sigmaAA-inc-upsilonx-dy}
\end{figure*}

Fig.~\ref{fig:sigmaAA-incoh-psix-dy} shows the differential cross section for the incoherent photoproduction of $\psi(1S)$ (left panel) and $\psi(2S)$ (right panel) as a function of 
the meson rapidity at $2.76$ TeV. These results are obtained using the GBW model and 
the wave functions in the potential approach and are compared, in the case of ground state 
production, with the single data point from ALICE collaboration \cite{Abbas:2013oua}. As can 
be seen, the results obtained with the formalism described above do not describe the data. 
This means that the approach has to be improved in the case of incoherent production, and 
a theoretical further analysis is necessary. Nevertheless, for completeness, in Fig.~\ref{fig:sigmaAA-inc-upsilonx-dy} we show the predictions for incoherent 
photoproduction of $\Upsilon (1S)$ (left) and $\Upsilon(2S)$ (right) at $5.02$ TeV obtained 
with the same formalism. As in Fig.~\ref{fig:sigmaAA-coh-upsilonx-dy}, there is a spread 
in the predictions obtained with different interquark potentials in the case of ground 
state photoproduction. Such a strong sensitivity to the potential models would, in principle,
enable one to set more stringent constraints on the heavy quark interaction potential
once the corresponding precision data become available.

Note, the inclusion of the gluon shadowing is necessary to describe the data points 
for $J/\Psi$ photoproduction in the nuclear target case. However, there is a question 
about the size of a possible double counting between the nuclear structure function and 
the gluon shadowing accounted for in our phenomenological approach. In order to better 
understand this point, in Fig.~\ref{fig:sigmaAA-coh-upsilonx-dy} we have compared the 
differential cross sections with and without gluon shadowing. In fact, the finite coherence 
length effects are relevant at large rapidities only, whereas the gluon shadowing corrections 
are important closer to mid-rapidity. We notice that, while the Glauber corrections are included
in all cases in Fig.~\ref{fig:sigmaAA-coh-upsilonx-dy}, apparently they are not sufficient 
for the data description, while the additional gluon shadowing represents a much 
bigger effect relevant to achieve such a description. Thus, if there is any double counting, 
we expect it to be relatively small and insignificant for our first analysis.
\begin{figure*}[!htbt]
\begin{minipage}{0.48\textwidth}
 \centerline{\includegraphics[width=1.0\textwidth]{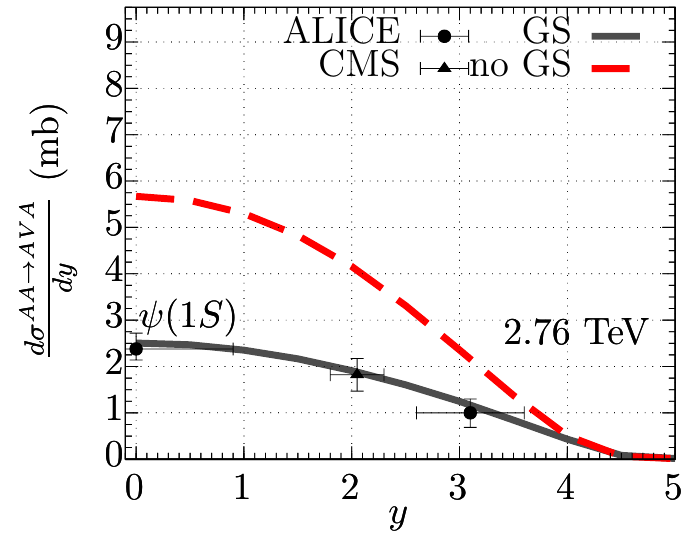}}
\end{minipage} \hfill
\begin{minipage}{0.48\textwidth}
 \centerline{\includegraphics[width=1.0\textwidth]{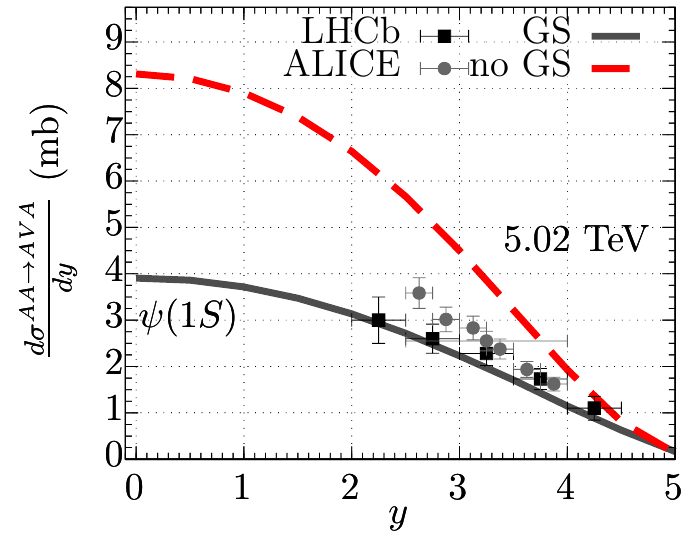}}
\end{minipage} \hfill
\caption{Comparison of the differential (in rapidity) cross sections of $J/\psi$ 
photoproduction with and without the gluon shadowing (GS) for BUT potential versus the experimental data 
in PbPb UPCs at $2.76$ TeV (left) and $5.02$ TeV (right).}
\label{fig:sigmaAA-coh-upsilonx-dy}
\end{figure*}

\section{Conclusions}
\label{Sect:Concl}

With the idea of reproducing what is already well known, we have described all the existing data 
on photoproduction of $J/\psi(1S)$, $\psi(2S)$, and $\Upsilon(1S)$ off the proton target 
with a good accuracy using the GWB and KST dipole models and different interquark interaction
potentials for the light-front vector meson wavefunctions. We have also made predictions for the 
$\Upsilon(2S)$ photoproduction in this case, for the sake of completeness.

After establishing that the proton target case was well understood in our case, we moved 
to the nucleus target. In order to do so, we included the important effects of gluon shadowing 
and finite coherence length of the quark-antiquark dipole. So far in the literature, 
the combined use of the gluon shadowing fitted to data and the interquark potential models 
(with Melosh spin rotation) for the quarkonia wavefunction has not been employed. Furthermore, 
no predictions for the bottomonia using the potential model were available as far as we know.

To start with, we considered the case of coherent photoproduction when the whole nucleus 
interacts with the photon. We achieved a good description of the $J/\psi(1S)$ and $\psi(2S)$ 
ALICE and CMS data sets from $AA$ UPCs with a center-of-mass energy of 2.76 TeV. At a higher 
energy of 5.02 TeV, our description is consistent with the LHCb data, while there is some 
tension with the ALICE data. We have also made predictions for the  $\Upsilon(1S)$ and 
$\Upsilon(2S)$ photoproduction in $AA$ UPCs that can potentially help in determining 
the best interquark potential for the $b\bar{b}$ vector mesons.

In the incoherent case, we make predictions for the same four vector meson states. 
The only available datapoint (from ALICE) does not agree with our calculation. Nevertheless, 
we chose to show these predictions because there is in the literature a well-known difficulty 
of describing the incoherent production data of any type. Therefore, we have demonstrated 
the best description that can be achieved with the (approximate) models described in this 
paper and in the literature.

As a final conclusion, we would like to mention that the coherent production case can be 
described very well if the complete treatment of nuclear effects and vector-meson light-front
wavefunctions discussed here is employed. Regarding the incoherent case, we see a strong need 
for further precise measurements and a better theoretical description of the underlying physics.

\section*{Acknowledgments}

This work was supported by Fapesc, INCT-FNA (464898/2014-5), and CNPq (Brazil) for CH, EGdO, and HT. 
This study was financed in part by the Coordenação de
Aperfeiçoamento de Pessoal de Nível Superior -- Brasil (CAPES) --
Finance Code 001. The work has been performed in 
the framework of COST Action CA15213 ``Theory of hot matter and relativistic heavy-ion collisions''
(THOR). R.P.~is supported in part by the Swedish Research Council grants, contract numbers
621-2013-4287 and 2016-05996, as well as by the European Research Council (ERC) under 
the European Union's Horizon 2020 research and innovation programme (grant agreement No 668679). 
This work was also supported in part by the Ministry of Education, Youth and Sports of the Czech
Republic, project LTT17018. EGdO would like to express a special thanks to the Mainz Institute for
Theoretical Physics (MITP) of the Cluster of Excellence PRISMA+ (Project ID 39083149) for its
hospitality and support.

    \bibliography{bib}

\end{document}